\documentclass[12pt,preprint]{aastex}
\begin{document}

\title{Distances to Anomalous X-ray Pulsars using Red Clump Stars}
\author{Martin Durant and Marten H. van Kerkwijk}
\affil{Department of Astronomy and Astrophysics, University of
  Toronto\\  60 St. George St, Toronto, ON\\ M5S 3H8, Canada }
\keywords{pulsars: anomalous X-ray pulsars, distance measurement, ISM,
reddening }

\begin{abstract}
We identify ``red clump stars'' -- core helium-burning giants -- among
2MASS stars and use them to measure the run of reddening with distance
in the direction of each of the Galactic Anomalous X-ray Pulsars
(AXP).  We combine this with extinction estimates from X-ray
spectroscopy to infer distances and find that the locations of all AXP
are consistent with being in Galactic spiral arms.  We also find that
the 2--10\,keV luminosities implied by our distances are remarkably
similar for all AXP, being all around
$\sim\!1.3\times10^{35}{\rm\,erg\,s^{-1}}$.  Furthermore, using our
distances to estimate effective black-body emitting radii, we find 
that the radii are tightly anti-correlated with pulsed fraction, and
somewhat less tightly anti-correlated with black-body temperature.  We
find no obvious relationship of any property with the dipole magnetic
field strength inferred from the spin-down rate.
\end{abstract}
\maketitle

\section{Introduction}
The Anomalous X-ray Pulsars (AXPs) are young, energetic, X-ray bright
isolated neutron stars, with spin periods of the order 10\,s.  They are
called {\em anomalous} since their luminosity far exceeds the energy
available from spin-down, and no binary companions are seen. AXPs
(along with the related Soft Gamma-ray Repeaters or SGRs) are now
believed to be {\em magnetars} (Thompson \& Duncan, 1996). Magnetars
have huge external magnetic fields ($\sim 10^{14}$G) and even larger
internal fields. It is the decay of the magnetic flux which provides
the luminosity seen, and is responsible for a whole array of
observational effects such as bursting and giant flares. See Woods \&
Thompson (2004) for a summary of recent observational data on
magnetars, and how they are modeled.

Since they are young remnants of massive, short-lived progenitors, all
of the AXPs 
are found in the Galactic plane (except for CXOU~J010043.1$-$721134
which is in the Small Magallanic Cloud). This causes a major obstacle
to observations: high interstellar extinction, manifested as
photo-electric edges from different elements in the soft X-ray band,
and as continuum extinction from dust in the optical and
near-infrared.  Since the amount of extinction has been difficult to
estimate accurately, the spectral energy distributions of AXP have
been subject to large uncertainties (Hulleman et al.\ 2004; Durant \&
van Kerkwijk 2005, 2006).

Furthermore, even if the interstellar absorption is
well-characterized, distances and therefore absolute fluxes are
difficult to determine.  The simplest distance estimate is made by
requiring that the black-body component typically inferred from the
X-ray spectrum arises from a neutron-star sized area.  We do not,
however, expect the surfaces of AXPs to be homogeneous, both on
observational grounds (they pulsate) and from theoretical
considerations (the magnetic field, which affects the heat conduction,
will vary across the surface).

For AXPs that are associated with supernova remnants or other
interstellar structure, a more direct distance estimate can be made
using 21\,cm \ion{H}{1} measurements and the Galactic rotation curve.
Convincing cases for associations with supernova remnants have been
made for two AXPs: 1E~2259+589 with CTB~109 (Gregory \& Fahlman,
1980), and 1E~1841$-$045 with Kes~73 (Sanbonmatsu \& Helfand 
1992).  Furthermore, Gaensler et al. (2005) found an \ion{H}{1} bubble
coincident with the direction of AXP 1E~1048.1$-$5937, which they
suggest was created by the winds of the massive progenitor of the AXP
(see Muno et al. 2006 for a discussion of possible massive progenitors
to AXPs).
Even for these systems, however, distance estimates can be rather
uncertain, with different authors finding inconsistent results (we
discuss this further in \S4 and~5).

For sources in the Galactic plane, a different clue to the distance is
available if one has a good measure of the interstellar extinction,
and the run of extinction with distance can be determined independently.  
This works because the extinction increases with distance, more so towards
star-forming regions and molecular clouds.

The so-called red clump method provides a means for deriving the
function of reddening versus distance in any given line of sight,
based on field stars over a relatively small area.  L\'opez-Corredoira
et al.\ (2002) noted that in an infrared color-magnitude diagramme of a
stellar cluster ($J-K$ versus $K$ for example), core helium-burning
giants, or red clump stars, form a well-defined and easily-identified
concentration of stars redward of the main sequence.  Stars spend up
to 10\% of their lifetime in this phase, and are much more luminous
than typical main sequence stars.  Because their helium cores all have
roughly the same mass, their luminosities are largely independent of
the total stellar mass.  Furthermore, their infrared colours are
insensitive to metallicity.  As a result, they are good infrared
standard candles and if the red clump can be identified at each of a
range of distances, then the reddening at each distance can be
calculated.  The method has been used not only by Lopez-Corredoira et
al.\ (2002) to measure the density distribution of stars in the
Galaxy, but also by, e.g., Drimmel et al.\ (2003) to map the
distribution of dust.

Here, we apply the red-clump method to measure distances to the
Anomalous X-ray Pulsars.  In section \S2, we first describe how we
applied the red-clump method, focussing on the AXP for which the
method was most tricky, and in section \S3 we use this field to test
the reliability of the method, and to estimate uncertainties in the
derived values.  In \S4, we present the results of the red clump
method applied to each of the Galactic AXPs, and we discuss the
implications in terms of distance. In \S5, we compare our new
distances with those in the literature, in particular for the
controversial case of 1E~2259+589 and the associated supernova remnant
CTB~109.  In \S6, we use our results to infer luminosities and
emitting radii, and discuss how these depend on other AXP properties.
We draw conclusions in \S7.

\section{Method}
Our method closely follows that described by L\'opez-Corredoira et
al.\ (2002).  We start with data from the 2MASS catalog (Skrutskie et al.,
2006) in the J- and K$_S$-bands. (From here on, we refer to the K$_S$ band as 
K for brevity.) Since all the Galactic AXPs lie along
the plane, we find a large number of 2MASS stars at relatively small
radii around each source.  For the analysis, we initially chose the
9999 nearest stars and used these to construct $J-K$ versus $K$
color-magnitude diagrammes (covering circular fields around each AXP
with radii in the range 9\farcm6 to 20.6\farcm).  Below, we first
discuss in some detail the results for the AXP 1E~1048.1$-$5937, for
which applying the method proved to be most tricky.

\begin{figure}
\begin{center}
\includegraphics[width=\hsize]{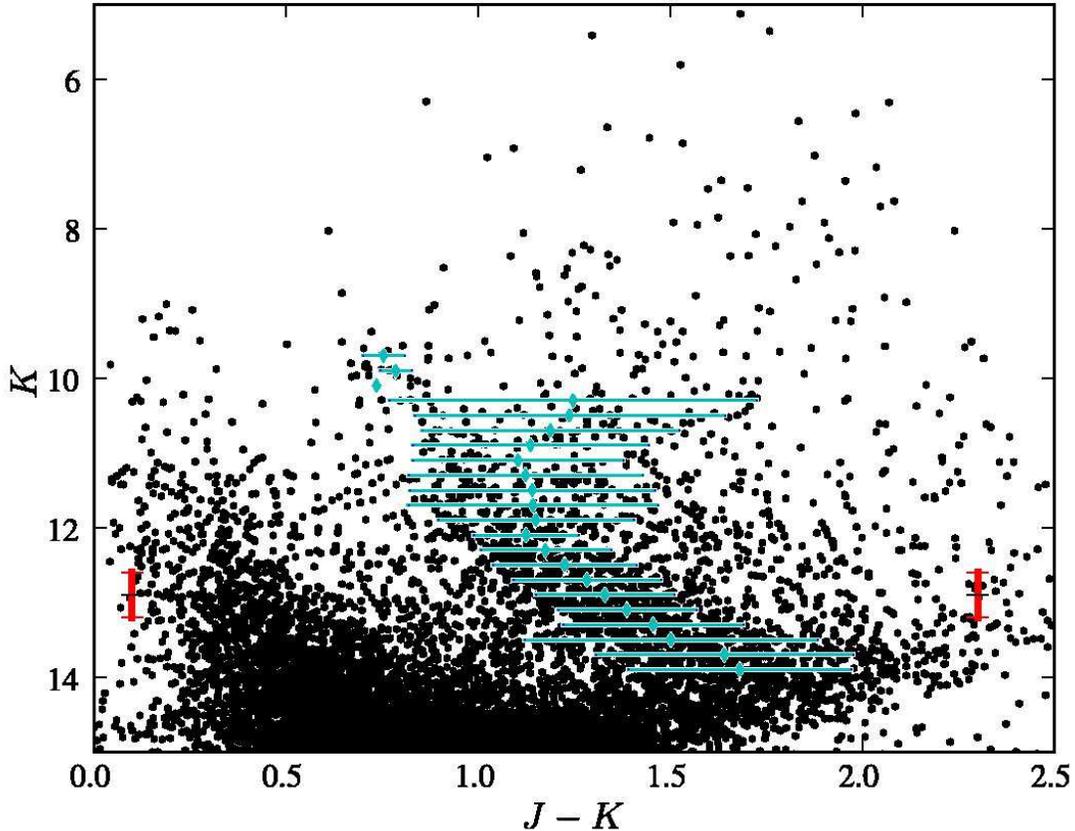}
\caption[0]{Color-magnitude diagram for 9999 stars within 11\farcm4 of
  1E~1048.1$-$5937. Main-sequence stars are found from about
  $(J-K,K)=(0.2,10)$ to (0.5,14), and red-clump stars show up as an
  over-density from about (1.0,10) to (1.7,14).  The fitted location of
  the red-clump peak (diamonds) and its extent (bars, showing
  $\pm1\sigma$ or roughly the full width at half maximum) are marked.
  The error bars (left and right edges) show the location and size of
  the strip used to create the histogram shown in Fig.~\ref{hist}.
  The poor fits near $K=11$ are discussed in
  \S3.}\label{CMD}\label{marked}
\end{center} 
\end{figure}

Figure \ref{CMD} shows the colour-magnitude diagramme for
1E~1048.1$-$5937.  On this graph, reddening alone causes stars to move
to the right and slightly down (redder and fainter) and increased
distance alone causes stars to move down (fainter in all bands).  From
the Figure, one sees that the main sequence (on the left of the
diagramme) contains the bulk of the stars. Due to the large range in
intrinsic luminosities, stars from different distances and reddenings
show up as an amorphous conglomeration, with an increasing range in
color towards the faint end.  The red clump, however, shows up as a
clearly defined stripe (center), representing stars of the same
luminosity and color at different distances and reddenings.  Since
red-clump stars are relatively rare, few are found at magnitudes
brighter than about $K=9$ (distance $\lesssim\!1.3\,$kpc).  
Stars redward of the red clump are either poorly measured (e.g.,
blended in the K band), background super-giants or young stars with
infrared excess.

\begin{figure}
\begin{center}
\includegraphics[width=\hsize]{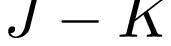}
\caption[0]{Histogram of the $J-K$ values of stars extracted from
  Fig.~\ref{CMD} in a 0.6\,mag interval around $K=12.9$. Overdrawn is
  a fit to this histogram using 
  Eq.~\ref{fitter}, with the curve showing the analytical function and
  the points the best-fit values. }\label{hist}
\end{center} 
\end{figure}

One can select any range in $K$ in the diagramme in order to find the
peak of the red-clump star distribution for that range.  Our
method works by identifying this peak for various ranges in K-band
magnitude.  As an example, we have selected a 0.6\,mag-wide strip around
$K=12.9$ for analysis.  The histogram of
$J-K$ colors for the selected stars is shown in Fig.~\ref{hist}. To
find the peak of the red-clump distribution, we fit a Gaussian
function to the histogram values. The main complication is
contamination by background, highly reddened main sequence stars (see
again Fig.~\ref{CMD}). These tend to skew the distribution, and for
$K$ approaching 14 overwhelm the Gaussian feature.  To account for these
contaminants, we fit the histogram with a power law plus Gaussian:
\begin{equation}
y = A_{\rm cont} (J-K)^{\alpha} + A_{\rm RC} \exp \left(-\frac{1}{2} \left[
  \frac{((J-K)-(J-K)_{\rm peak})^2}{\sigma^2}\right] \right) \label{fitter}
\end{equation}
where $(J-K)$ is the stellar color, $(J-K)_{\rm peak}$ is the peak of the
distribution, $\sigma$ the width of the peak, $A_{\rm RC}$ and $A_{\rm
cont}$ the normalizations of the red-clump and contaminant terms, and
$\alpha$ the power-law index.
\clearpage

\clearpage
Equation \ref{fitter} is fitted using a standard least-squares
algorithm, assuming the errors in each histogram bin are the same.
This process is repeated for successive bins in $K$ until the 2MASS
completeness limit is reached. Figure \ref{marked} shows the fitted
red clump positions overdrawn on the color-magnitude
diagramme. Features from this plot, such as the poor fits around
$K=11$, will be discussed in the next section.

Since our algorithm finds only local minima in the $\chi^2$ surface,
it is necessary to supply 
reasonable starting values to the routine. We found that if we chose
our strips in $K$ to overlap, and used the best-fit values of the
previous histogram, smooth changes in the peak were well followed. To
increase robustness, our algorithm also checks the relative residuals
of fits performed with a few different initial values of $(J-K)_{\rm
peak}$; this covers the case where there is a sudden jump in
reddening. Note that the choice of the range in $K$ of each histogram
and the extent to which these strips overlap is arbitrary, but should
not change the results. We test this below.

Assuming the intrinsic color $(J-K)_0=0.75$ and the intrinsic
luminosity $M_K=-1.65$ (Wainscoat \& Cowie 1992), the translation of
$J-K$ versus $K$ into reddening versus distance corresponds to
decomposing the reddening and distance vectors on the diagramme for
each red-clump peak found. The infrared color excess can be expressed
as a visual reddening through (e.g., Schlegel et al.\ 1998):
\begin{equation}
A_V = \frac{(J-K)_{\rm peak} - (J-K)_0}{0.164}  \label{redconv}
\end{equation}
The distance is then (correcting for reddening in K):
\begin{equation}
d = 10^{0.2(K-M_K-0.112A_V)}\times10\,\rm{pc}.\label{distance}
\end{equation}

\section{Robustness and Uncertainties}\label{varify}
The approach discussed above for converting a 2MASS dataset into a run
of reddening with distance does not yet yield estimates of the
uncertainties.  Here, we investigate the error analysis for the field
of 1E~1048.1$-$4937 in detail.  We chose this field since it
exemplifies the problems faced, and lies near one of the most
complicated regions of the Galaxy, the Carina Complex. None of the
other five AXP fields are more complex.

For our discussion, we estimate the formal uncertainties on the
location of the red-clump peak with $\sigma/\sqrt{N_{\rm RC}}$ (where
the number of red-clump stars $N_{\rm RC}=\sqrt{\pi}\sigma A_{\rm
RC}$, with $\sigma$ and $A_{\rm RC}$ from Eq.~\ref{fitter}), which
should be a good approximation if the Gaussian is a reasonable fit for
the distribution of infrared colors in the red clump, and if the
contamination by other stars is not too large.  For the test field,
the FWHM of the peak is typically $\approx0.3$\,mag and the number of
stars of order 100s, giving uncertainties
of the position of the peak $\sim0.02$\,mag (one-sigma).

In principle, one
could derive better formal uncertainties, e.g., by calculating the
$\chi^2$ values around the best-fit function, or -- more appropriately
for a counting problem with Poissonian statistics -- using a
maximum-likelihood method (which might even take into account the
probability distributions in $J$ and $K$ for each star).  We believe,
however, that a better approach is to keep the estimate of the formal
errors simple and combine it with empirical tests for systematic
effects using different subsets of stars and using multiple fields.

The first test we did was to try different cuts on the 2MASS data
using the magnitude error estimates and photometric quality flags, in
order to keep only the best measured points.  We found, however, that
this had the sole effect of truncating the CMD at brighter magnitudes,
without significantly tightening the spread in the red-clump stars at
any magnitude.  This suggests that any spread in the red-clump stripe
on a CMD beyond the typical measurement error is due predominantly to
field inhomogeneities.  Thus, in order to find the reddening to as
large a distance as possible, we continue to use all the available
2MASS stars in a given field.

An important feature that can be seen for the bins between $K=10.5$
and 11.5 in Fig.~\ref{marked} is that there seems to be overdensities
at both $J-K=0.9$ and $J-K=1.2$ and that the fitted points lie in
between these two.  This suggests that the reddening at these
distances is variable across the field. Since two peaks are visible,
part of the field might be obscured by denser interstellar matter.

\begin{figure}
\begin{center}
\includegraphics[width=\hsize]{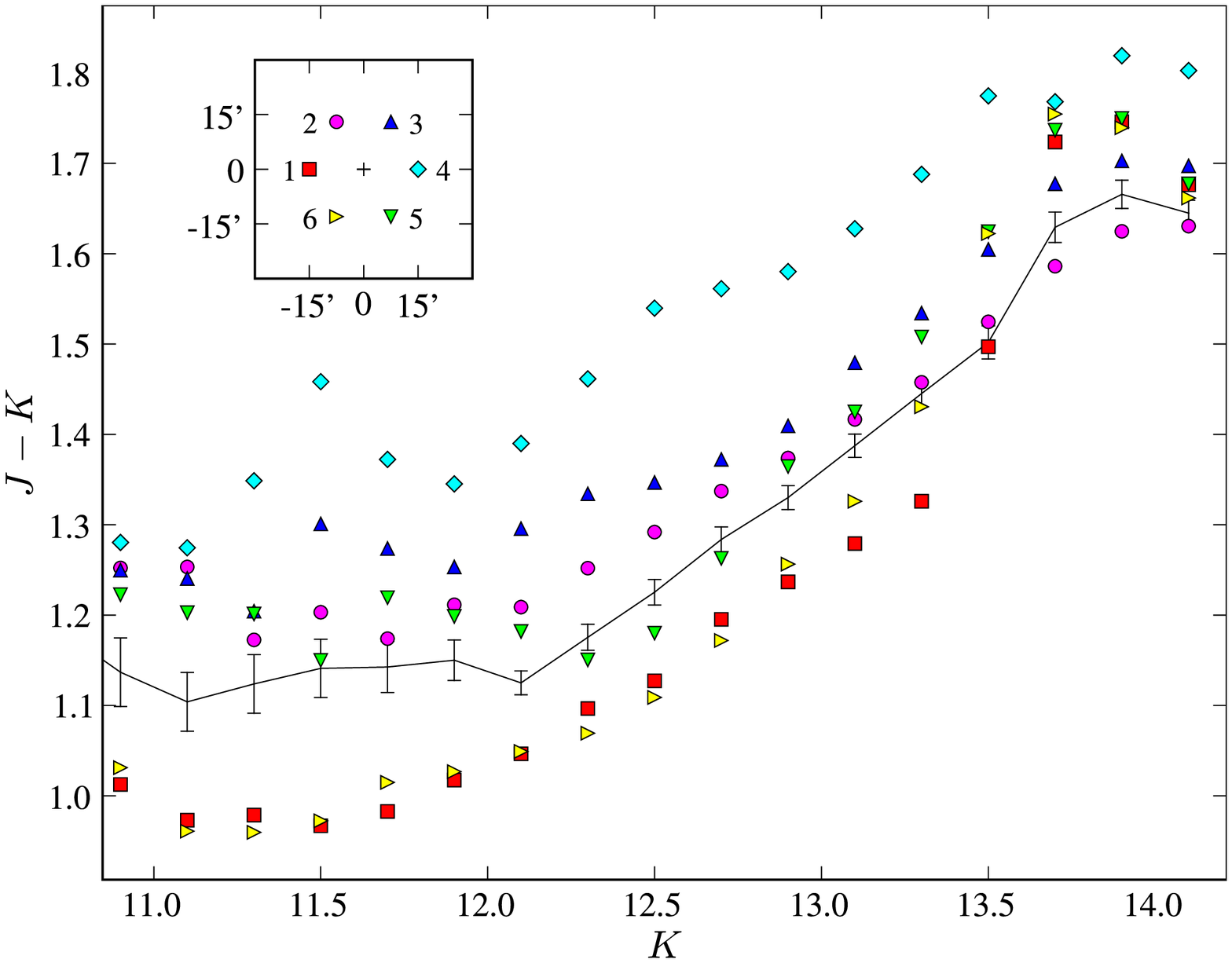}
\caption[0]{Color of the red-clump stars as a function of magnitude
  for the field around 1E~1048.1$-$5937 (curve with error bars) and
  six fields surrounding it (symbols).  Field~1 is offset by
  15\arcmin\ to the East, and the other fields are placed
  anti-clockwise with increasing number (see inset). }\label{circle}
\end{center} 
\end{figure}

To investigate the inhomogeneity in reddening, and to determine how
this affects the sensitivity of method, we analyzed six fields at a
distance of 15\arcmin\ from 1E~1048.1$-$5937 (note that these fields
overlap somewhat with the target field).  Figure \ref{circle} shows
the resulting graphs of $J-K$ versus $K$ (we show this graph rather
than $A_V$ versus $d$, since $d$ depends on both values).  One sees
that the distribution of reddenings is trimodal, with fields 1 and 6
showing consistently less reddening as a function of distance than
fields 2, 3 and 5, but merging to the same value for faint $K$ (note
that some fields appear to show a decrease in reddening; this is due to
the same bimodality that affects the central pointing). Field 4 shows
consistently higher reddening, and is the line of sight passing
towards the cluster to the west.  Thus, there
is structure in the Galactic dust distribution: a dust cloud appears
to cover the field gradually from the North-West with increasing
distance.  Considering the complex structure of the local Carina
Nebula, this is not surprising; it is simply a consequence of the
large reddening gradients in the area.

\begin{figure} 
\begin{center}
\includegraphics[width=\hsize]{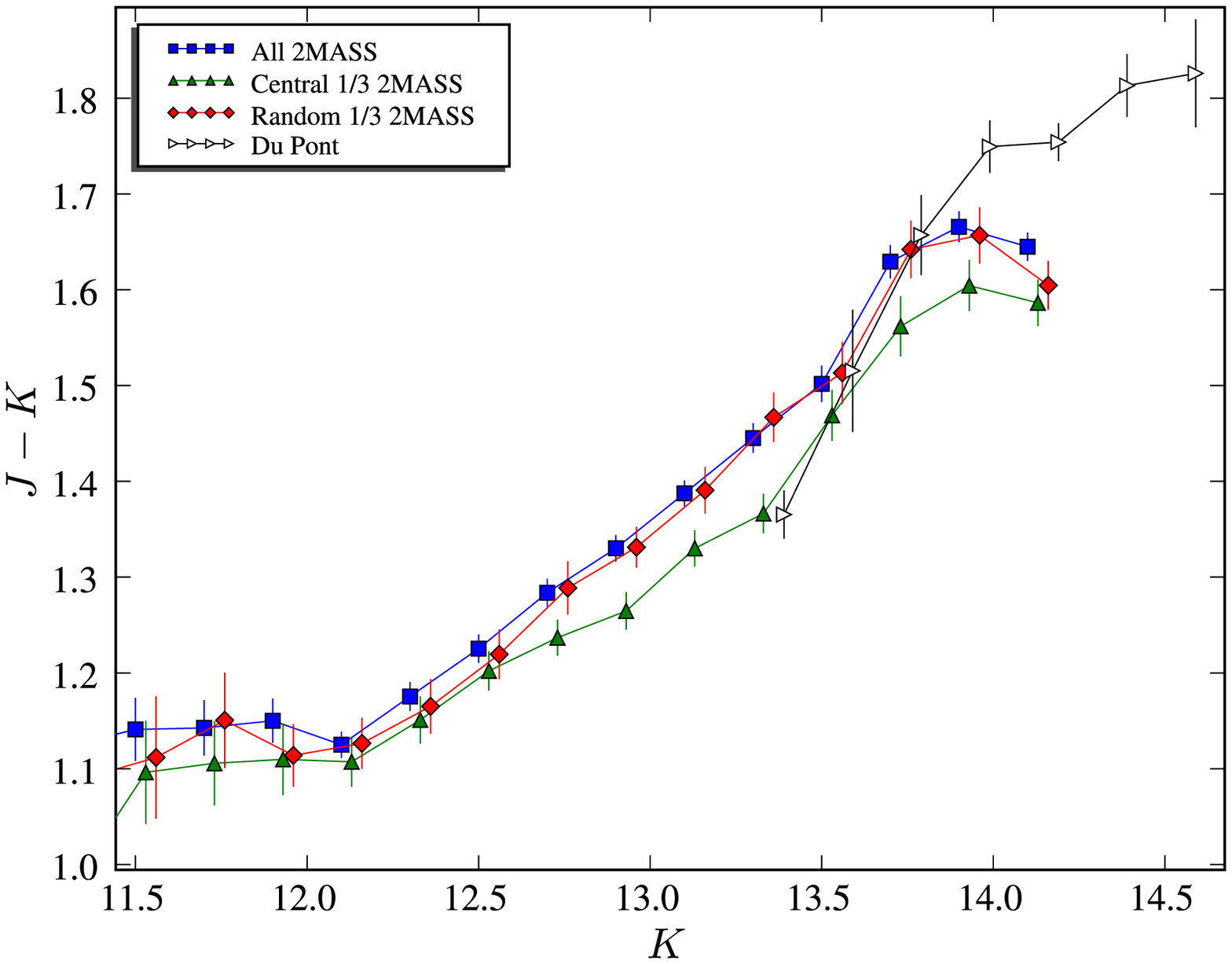}
\caption[0]{Color of the red-clump stars versus magnitude in the
  direction of 1E~1048.1$-$5937 for different subsamples (see inset).
  The curves have been slightly offset in $K$ for
  clarity. }\label{central} 
\end{center} 
\end{figure}

From Fig.~\ref{circle}, comparing fields with similar runs of $J-K$ as
a function of $K$ (such as 1 and 6, and 2 and 3), one sees that the
uncertainties inferred from the Gaussian fit (which are similar for
all fields) agree roughly with the scatter between those fields,
suggesting that they are reasonable estimates of the uncertainties.
(One also sees that compared to the scatter along the curve, the error
bars appear to be overestimates -- this, however, is a consequence of
the fact that the points are not independent, as the strips in $K$
overlap.)
\clearpage
\clearpage
Given the inhomogeneity in reddening, we checked whether chosing a
smaller field would yield a tighter fit.  Figure~\ref{central} shows a
comparison of the results for the original set of 9999 stars around
1E~1048.1$-$5937 with the results for just the central one-third of
those stars, as well as, for comparison, for a randomly chosen
one-third of the stars.  Interesting behavior is seen in the
randomly-chosen curve: generally it follows the curve of the original
sample (with increased error bars), but around $K=12$ the inferred
$(J-K)$ color decreases a little and is lower than inferred from the
original sample.  This is due to the competition between the two areas
with different reddening in the field.  With fewer stars, the fit is
more liable to jump from one over-density to the other.  For the
central sub-sample, however, this does not occur.  Instead, it shows
consistently lower reddening than the original sample at all values of
$K$.  The errors are only slightly larger, since the smaller number of
stars is offset by the smaller spread in the colors of the red-clump
stars.

From our tests, we conclude that in the field of 1E~1048.1$-$5937 a
dark cloud towards the North obscures some fraction of area within the
11\farcm4 radius covered by the original 9999 stars,
but it does not cover the source and hence should be ignored in
determining the run of reddening with distance.

More generally, we conclude that, since our aim is not to determine
accurately the {\em average} reddening in the field, but rather to
find the reddening at the center of the field only, we should use the
minimum number of stars required for our method to work reliably.
Empirically, we find that this is $\sim\!3000$ stars; with fewer
stars, the red-clump peak becomes poorly defined due to small-number
statistics.

In the case that there remains substantial inhomogeneity within the
small field under examination (this is certainly possible, but it
turns out not the case for the AXP fields), we must raise an important
caveat to the use of the method. Because the over-desity finding
rutine picks the best Gaussian feature out of the color histograms,
it will be biased towards the more homogenous part of the field. If
the field is highly inhomogenous, it may not be able to locate the
red-clump satisfactoraly; if a compact dense cloud obscures part of
the field, then the method will give the reddening of the unobscured
part of the field, since the cloud will have strong column density
gradients across it.

\clearpage
\clearpage
The next effect we checked is that of the choice of the size and
overlap of the bins in $K$. In order to have as high a distance
resolution as possible, one would like to use bins in $K$ that are as
small as possible.  In order for the red-clump peak to be better
defined on each histogram, however, one would prefer more stars and a
narrower distribution of red-clump star colors. As discussed above,
increasing the field size in order to use more stars does not help
since the reddening may be non-uniform. The same is true for
increasing the K-bin sizes: different colors from different distances
will tend to blur the red-clump peak. Since in this case this also
comes at the cost of reduced distance resolution, we opt for bin sizes
as small as possible which still give reliable results: 0.6\,mag, with
steps of 0.2\,mag. The overlap factor does not change the results or
our method, but it is convenient computationally and also allows us to
determine slightly more accurately the locations of jumps in
reddening.
\clearpage
\begin{figure}
\begin{center}
\includegraphics[width=\hsize]{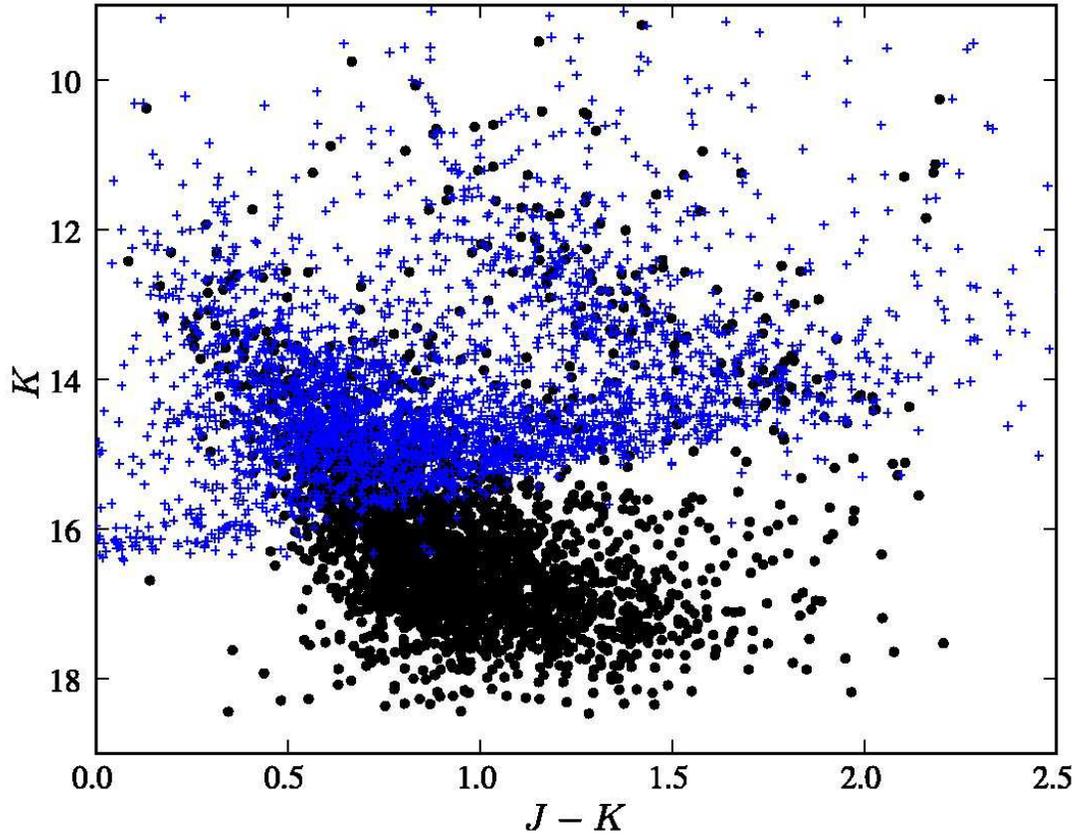}
\caption[0]{Deeper color-magnitude diagram of the field of
  1E~1048.1$-$5937, from Du Pont imaging. The Du Pont sources are
  shown as black circles, and the 2MASS points (from the closest 3300
  stars) as crosses.}
  \label{dupont}
\end{center} 
\end{figure}
\clearpage
The final thing we checked is the effect of the 2MASS limiting
magnitudes on our measurements.
Our precise estimate of the distance of 1E~1048.1$-$5937 from 2MASS
data alone is 
uncertain, because the last point in our graph of $J-K$ versus $K$  is likely
affected by the incompleteness to fainter and to redder sources. To
check for this, we obtained further, deeper JK infrared imaging data from
the Wide-field Infrared Camera on the Du Pont 2.5\,m telescope, Las
Campanas (see Persson et al. 2002). We subtracted a sky frame from
each image using the median of the science frames (this is required
because of an additive component to the noise from fringing),
registered and stacked the images. We tied the photometry directly to
2MASS using many stars cross-identified on each of the chips.

Figure \ref{dupont} shows the new, deeper color-magnitude diagram for
stars in the field of 1E~1048.1$-$5937. This includes stars from the
chip containing the AXP and two adjacent chips (each chip has a size
200\arcsec\ square). 
One sees that down to $K=13$, $J-K=1.4$, which corresponds to the last           
reliable red-clump location from the 2MASS data, the Du Pont                     
and 2MASS data are consistent, having overdensities at the same                  
colour.  However, for the next group, at $K=14$, $J-K=1.8$, the Du               
Pont data indicate a location redder than the 2MASS data would have              
suggested. Thus we conclude that the reddening rises more quickly
in this region, and the AXP is closer than otherwise would have
seemed. We show the $J-K$ values obtained from the Du Pont data in
Figure \ref{central} and use these for the final point on the
reddening curve for 1E 1048.1$-$5937 in Figure \ref{output}.

As mentioned above, the challenges faced in the field of
1E~1048.1$-$5937 are more severe than those for any of the other
fields; none show reddening gradients as large, and for all the other
sources (except 1E~2259+586, see \S\ref{results}) the appropriate value
of the color excess lies either well before or beyond the
completeness limit.  Using what
we have learned here, we perform our analysis on the central 3300
stars near each AXP, and infer uncertainties using the simple estimate
from counting statistics.

\section{Results}\label{results}
\clearpage
\begin{figure}
\begin{center}
\includegraphics[width=0.49\hsize]{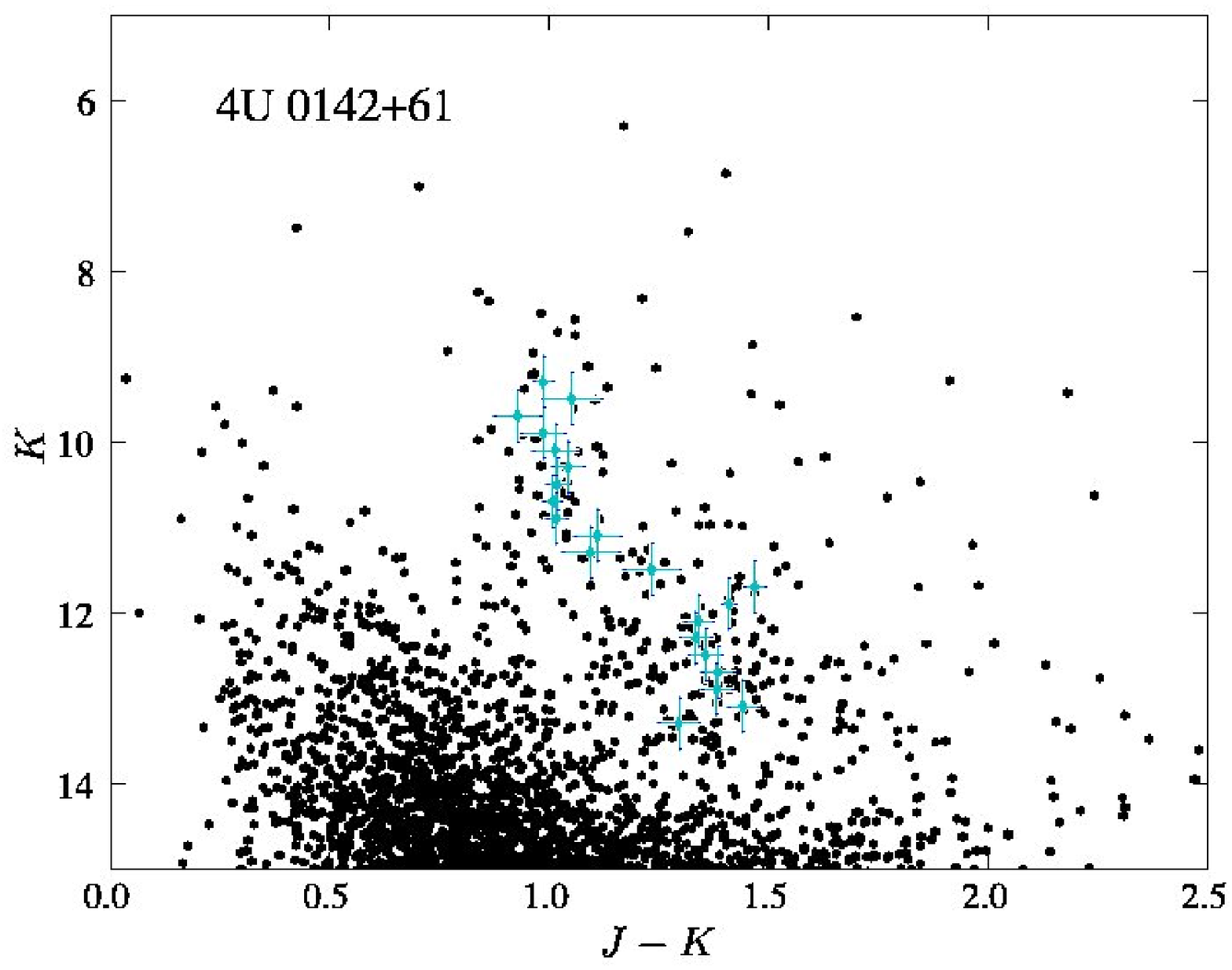}
\includegraphics[width=0.49\hsize]{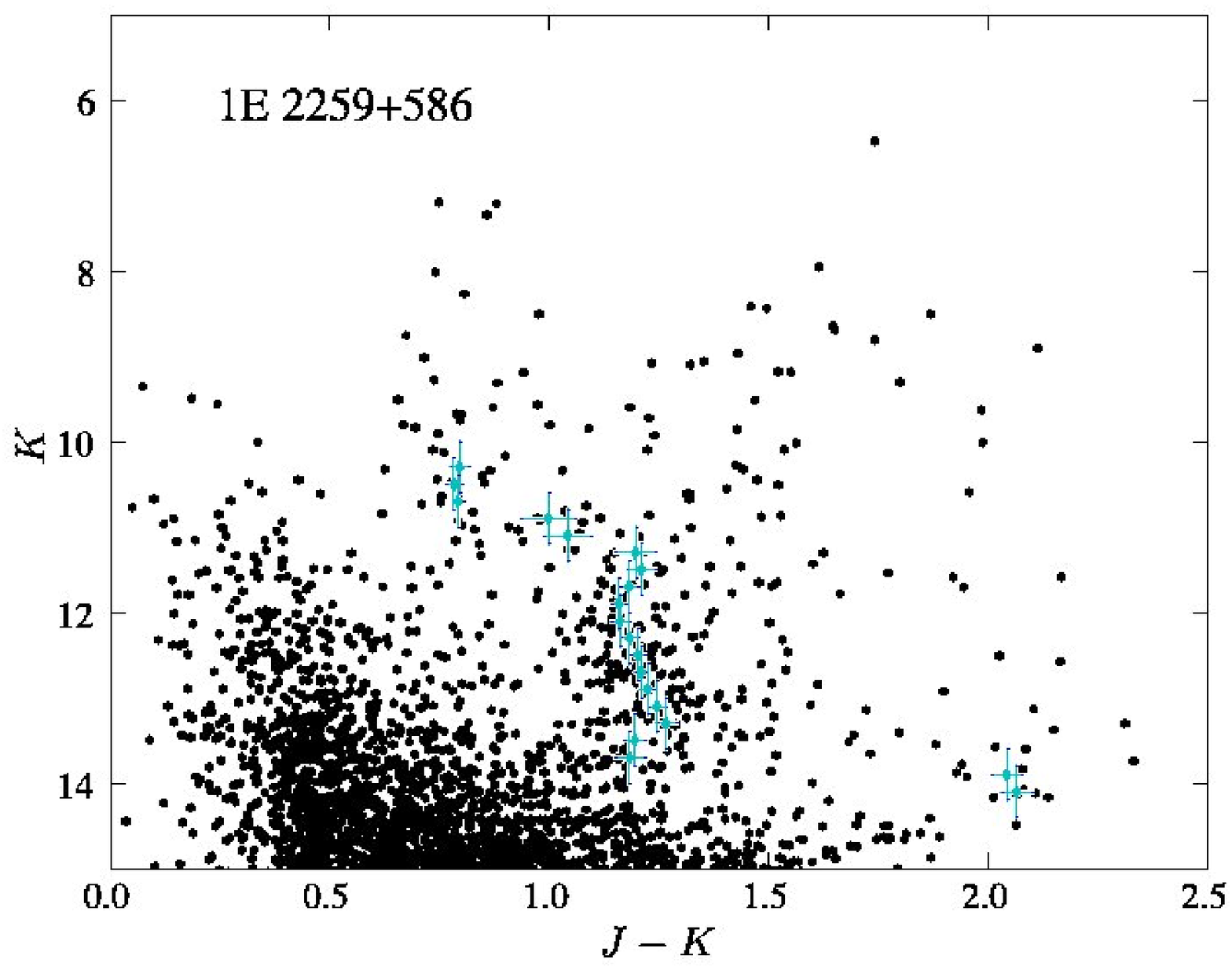}
\includegraphics[width=0.49\hsize]{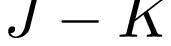}
\includegraphics[width=0.49\hsize]{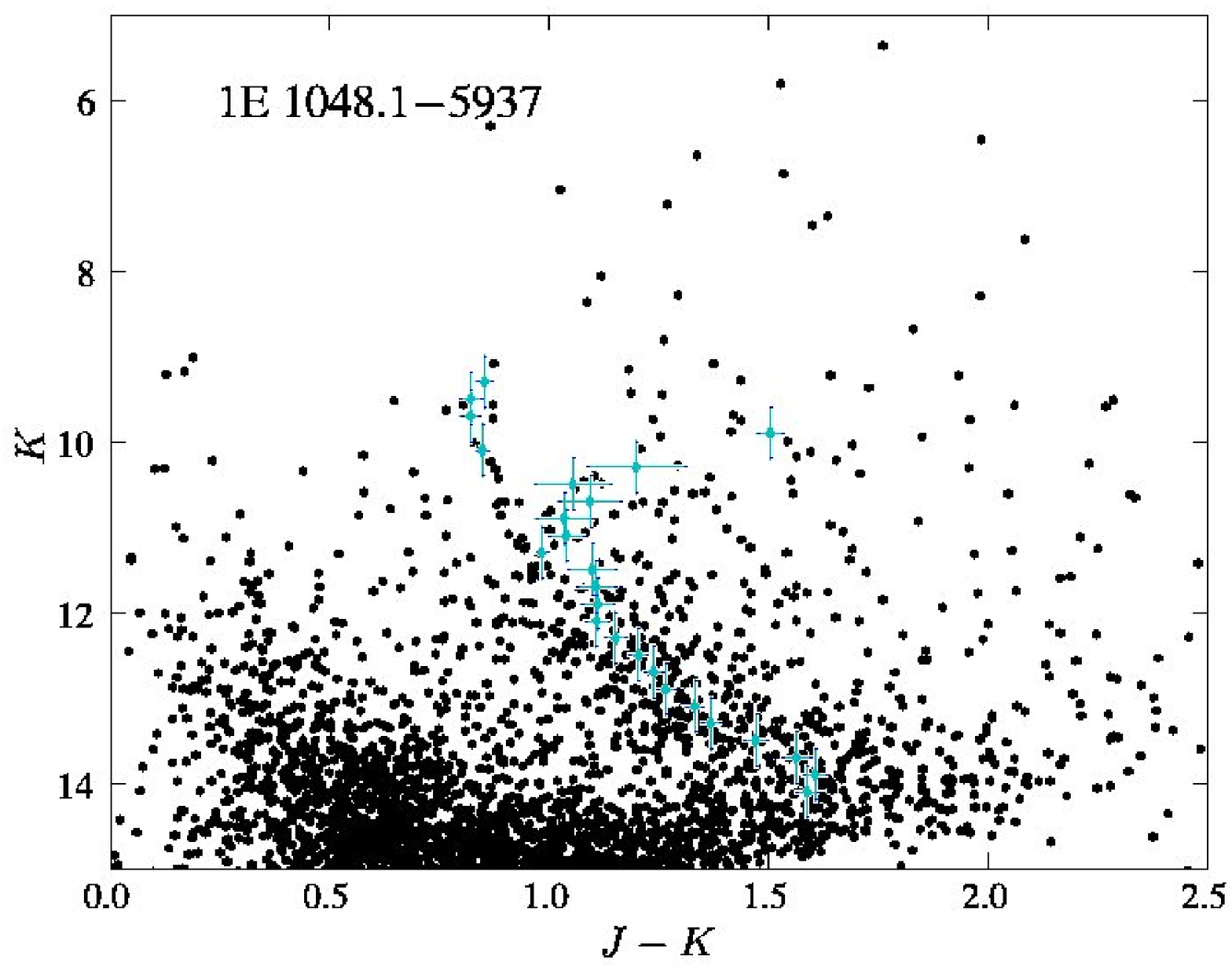}
\includegraphics[width=0.49\hsize]{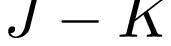}
\includegraphics[width=0.49\hsize]{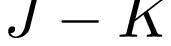}
\caption[0]{The color-magnitude diagrams used to infer the location
  of the red clump as a function of brightness, derived from the
  nearest 3300 2MASS stars around each object (within 5 to 12\arcmin).
  Overdrawn are the fitted red-clump peak colors together with the
  1-$\sigma$ uncertainties.}\label{CMDs}
\end{center} 
\end{figure}
\clearpage
We determined the color of the red clump as a function of brightness
using the nearest 3300 stars for each of the six Galactic AXP.  In
Table~\ref{fields}, we list the size of the field covered by these
stars, and in Fig.~\ref{CMDs} we show the corresponding
color-magnitude diagrams, with the fitted locations of the red-clump
stars indicated.

In converting our results to reddening as a function of distance, we
prefer to use non-overlapping points.  In choosing the points to use,
however, one has a certain freedom to pick those that have with the
smallest errors, and that show features the clearest.  We attempt to
do this, and check against the original color-magnitude diagrammes
that the breaks occur at the correct values of~$K$.  In interpreting
the fitted points, we also use the fact that reddening can only
increase with distance (e.g., we reject a poorly measured point with
high reddening in favor of a later, better measured point with lower
reddening).

\begin{figure}
\begin{center}
\includegraphics[width=0.49\hsize]{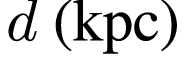}
\includegraphics[width=0.49\hsize]{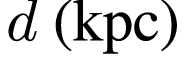}
\includegraphics[width=0.49\hsize]{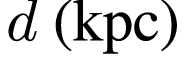}
\includegraphics[width=0.49\hsize]{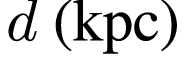}
\includegraphics[width=0.49\hsize]{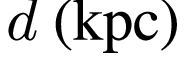}
\includegraphics[width=0.49\hsize]{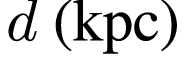}
\caption[0]{Run of reddening with distance along the line of sight to
  the six Galactic AXPs.  The best estimate for the reddening for each
  object is shown by a dashed horizontal line (for uncertainties, see
  Table~\ref{distances} and the text); note that the estimated $A_V=14$
  for 1E~1841$-$045 falls outside the graph. The grey point in the
  graph for 1E~1048.1$-$5937 is discussed in the text.}\label{output}
\end{center} 
\end{figure}

Figure \ref{output} shows the resulting runs of reddening with
distance, with associated uncertainties.  With X-ray extinction estimates for
the AXPs, distances can simply be read off the graphs by first
converting to $A_V$ using $A_V=N_H\times5.6\times10^{-22}$\,cm$^2$
(Predehl \& Schmitt, 1995).  For three of
the AXPs -- 4U~0142+61, 1E~2259+589  and
1RXS~J170849.0$-$400910 -- new, model-independent estimates of the
extinction were found by Durant \& van Kerkwijk (2006) using
photo-electric absorption edges in high-resolution X-ray spectra.
These are listed in Table~\ref{fields}; the uncertainties reflect the
accuracy with which the depths of the various features could be
measured. 
For one further AXP, 1E~1048.1$-$5937, our measurement was not very              
constraining, while for the remaining two, 1E~1841$-$045 and                     
XTE~J1810$-$197, no suitable high-resolution X-ray spectra were                  
available.  Hence, for these three AXPs, we use the extinction                   
estimates compiled in Woods \& Thompson (2004).                       
These estimates are based on broad-band fits to
the X-ray spectra, which means that their accuracy depends on the
adequacy of the model (typically a two-component model composed of a
power law and a black body).  From a comparison of similar estimates
with our direct measurements for the three AXPs, we expect that, at the
relatively high extinctions found for these two sources, the estimates
should be accurate to about~10\%.

Before continuing, we should address a possible worry: that a              
large uncertainty is introduced in the various conversions used.                 
First, above we implicity converted photo-electric absorption edges by           
metals to an equivalent hydrogen column $N_H$, and that in turn to               
visual extinction $A_V$.  This is not as uncertain as it appears,                
however, since the relation of Predehl \& Schmitt (1995) between X-ray           
and visual extinction, while quoted in terms of hydrogen column $N_H$,           
was based on X-ray measurements that were sensitive to metals along              
the line of sight, not hydrogen, just like ours.  Hence, there is no             
uncertainty associated with, e.g., the metallicity or the hydrogen to            
dust ratio.  A second worry one might have is about selective                    
extinction, i.e., variations in the interstellar extinction law                  
(usually parametrised by $R_V\equiv A_V/E_{B-V}$).  There is, however,           
far less variation in the infrared (e.g., Mathis 1990).  Since the               
optical extinction estimates of Predehl \& Schmitt were largely based            
on a standard extinction law, we should thus circumvent any problems             
at visual wavelengths by using the same standard relation between                
infrared and visual extinction.  (Indeed, it might well be that one              
would find a tighter relation than that of Predehl \& Schmitt (2005)             
if one redid the analysis in terms of infrared extinction.)

From Fig.~\ref{output}, one sees that for most sources, the measured
reddening places the source at a jump in the run of reddening with
distance.  As a result, the uncertainty in the distance estimate is
dominated not by uncertainty in $A_V$, but by the accuracy with which
the jump can be located, which in 
turn is determined by the size of the bins in $K$. A bin size of 0.6\,mag 
corresponds to a 15\% uncertainty in distance.  (As discussed
above, the size of the bins was chosen to give the best distance
resolution yet yield reliable measures of reddening without using a
large field and thus increasing the risk of bias by spatial variations
in reddening.)  The only source not located at a jump in reddening is
1E~1048.1$-$5937. Hence for this source,
uncertainty in the distance is dominated by the (relatively large)
uncertainty in the reddening estimate.
\clearpage

\begin{figure}
\begin{center}
\includegraphics[width=\hsize]{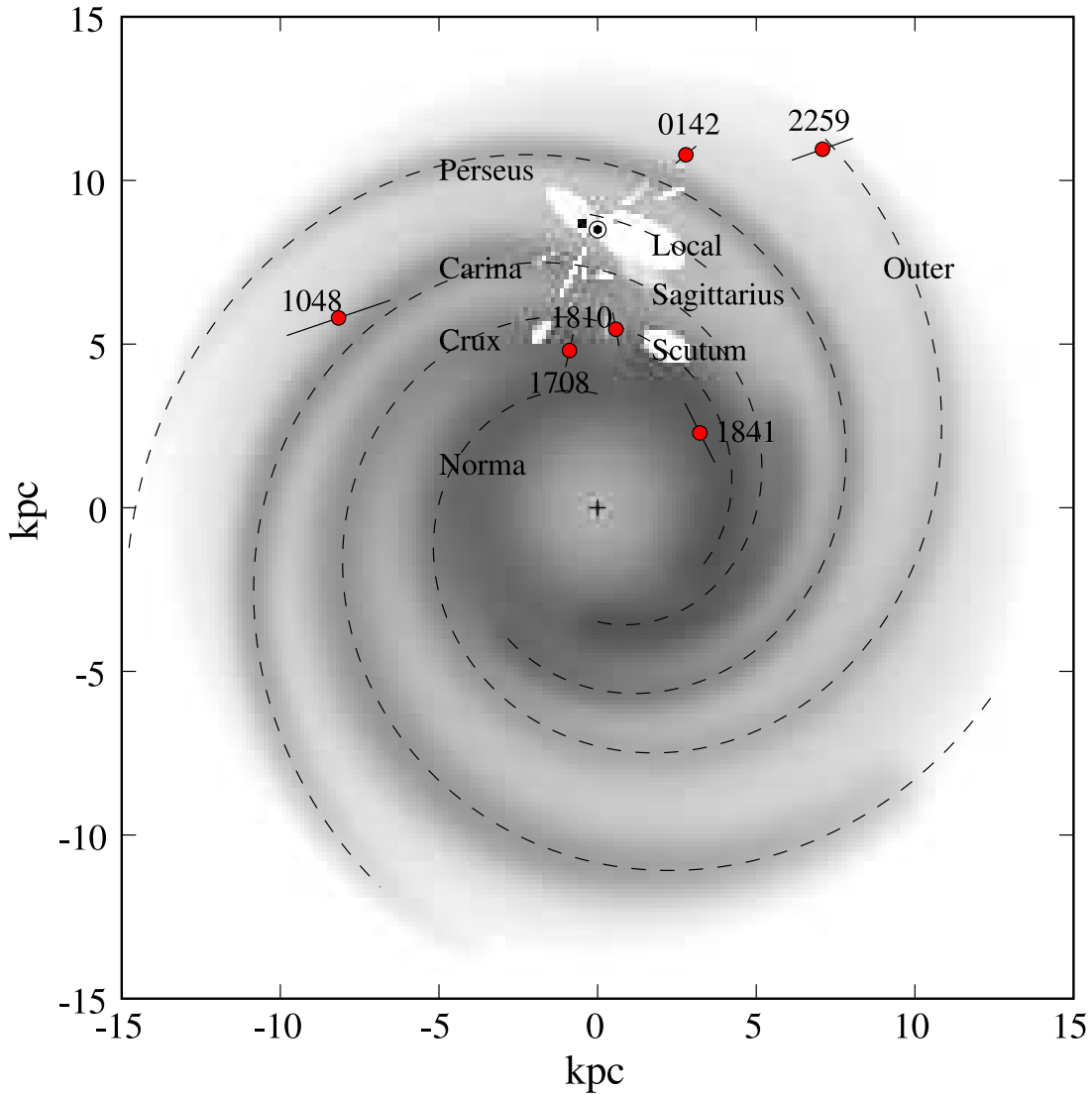}
\caption[0]{Schematic map showing the positions of the AXPs in the
  Galactic Plane relative to the Galaxy's spiral arms.  The AXPs are
  indicated with circles and labeled with abbreviated names, the
  Galactic center is marked with a ``$+$'' sign, and the location of
  the sun with a $\odot$ symbol (at a galactocentric distance of
  8.5\,kpc).  Spiral arms are shown using a fit with a simple four-arm
  logarithmic spiral model (dashed lines, see Cordes \& Lazio, 2002)
  and free electron density 
  (grey scale, taken from Cordes \& Lazio, 2002). The white patches are local
  interstellar cavities and the dark spot near the sun is the Gum
  Nebula. The grey scale backdrop has been rescaled to fit these axes.} 
  \label{map}
\end{center} 
\end{figure}
\clearpage
Table \ref{distances} gives final inferred distances and limits.
Below, we describe the results for each object in more detail, trying
to match rapid increases in reddening with known spiral arms, and
locating the AXPs in the Galaxy (see Fig.~\ref{map}). This can be
viewed as an update of the similar map shown in Gotthelf \& Vasisht
(1998; their Figure~4).  We focus on the 
outcome of our analysis, deferring a more detailed comparison of the
implied distances with previous determinations to \S\ref{consistency},
and a discussion of the implications in terms of luminosities and
other derived quantities to \S6.

\paragraph{4U~0142+61.}  In this field, two regions of reddening are seen,
$A_V\simeq2$ in the near field and $A_V\simeq4$ out to large
distances.  The latter confirms the statement by Hulleman et al.\ (2004)
that the reddening could not exceed $A_V\simeq5$ in this direction,
based on the colors of a background galaxy and of field stars.  From
the estimated reddening to the source of $A_V=3.6\pm0.2$, we conclude
that the AXP is located in a region of rapidly rising reddening at
$d\simeq3.5\,$kpc.  Likely, this region is associated with the Perseus
arm.

Looking in more detail at Fig.~\ref{CMDs}, one sees that for $K>13.5$,
one can no longer identify the over-density of red-clump stars.  The
stripe seen at lower $K$ does not continue and no over-density is seen
redward of this stripe either, as could be caused by a sudden increase
in reddening.  Most likely, this indicates simply that there are not
many red-clump stars beyond a distance of $\sim\!8\,$kpc.  Consistent
with this, one sees that the red side of the main sequence has a
smooth distribution as $K$ increases from 13 to 15; if the absence of
red-clump stars beyond $K>13.5$ were due to a large, sharp increase in
reddening, this should be visible for the main-sequence stars too.

\paragraph{1E~1048.1$-$5937.} Reddening rises continuously in this
field, almost linearly with distance out to $\sim\!12\,$kpc.  This is
probably due to the Carina Arm, which lies along the line of sight in
this direction (Dame at al.\ 2001; Fig.~\ref{map}).  Our estimate of
the reddening to the AXP places it at the tail end of this function.
The 2MASS completeness limit in the J-band causes
the stars to go undetected beyond a line $((J-K),K)=(1.5,15)$ to
(2.5,12). Our distance estimate is therefore based on the Du Pont
photometry we obtained, which goes deeper than the 2MASS survey (see
\S\ref{varify} and Figure \ref{dupont}). This
affects only the last point in the graph of $A_V(d)$. We include in
Figure \ref{output} the furthest point which would have been obtained
from the 2MASS data alone in faint grey. 

Despite the existance of a new estimate of reddening from Durant \&
van Kerkwijk (2006), the value given was not very constraining. We
therefore have used the old estimate base on fitting the sum of a
black-body and power-law to the X-ray spectrum. In this particular
case, that approach should be the least error prone, since the
power-law component is the shallowest of the AXPs, and hence has
relatively little influence on the low-energy region of the spectrum,
to which $N_H$ is most sensative (Woods \& Thompson,
2004). 

Our estimated distance is around 8.6\,kpc.
This distance estimate is inconsistent with that
of Gaensler at al.\ (2005); we return to this in \S\ref{consistency}.

\paragraph{1E~2259+589.} In this field, the reddening is low up to
$d\simeq3\,$kpc, next rises to $A_V\simeq3$, and then stays at that
value out to about 6.5\,kpc, where it shows a second and larger jump.
From Fig.~\ref{CMDs}, one sees that this second region of reddening is
inferred from only a few stars at around $J-K=2.0$ and $K=14$.  We
believe it is genuine, however, for several reasons.  First, at $K=14$
there is a notable absence of stars at $J-K=1.2$, below the location
of the stripe for brighter magnitudes.  Second, in contrast to what we
saw for 4U~0142+61 above, the main sequence's red edge does appear to
show a jump, from about $J-K=0.7$ at $K=13.5$ to $J-K=1.2$ at $K=14$.
Third, the two jumps have a nice correspondance with expected jumps
due to spiral arms, with the nearer one being due to the Perseus arm
and the farther one due to the Outer arm (Fig.~\ref{map}).

Assuming the second jump in reddening is real, the AXP's estimated
reddening of $A_V=6.3\pm0.7$ places it into this region, and thus
likely in the Outer arm, at $d\simeq7.5\,$kpc. Were the last point in
the curve of $A_V(d)$ affected by incompleteness of the 2MASS sample,
as was the case for 1E~1048.1$-$5937 (above), this would move the
point even further redward, make the jump steeper, and not
significantly affect the estimated distance.
In \S\ref{consistency}, we compare this with other distance estimates,
in particular of CTB~109, the well-studied supernova remnant
associated with 1E~2259+589. 

\paragraph{1RXS J170849.0$-$400910.} For this field, strong jumps in
the reddening curve are seen, which likely are associated with the
Carina and Crux spiral arms.  The estimated reddening for the AXP
places it into the second of the jumps, at about 3.5\,kpc.  From this
field, one also sees that the red-clump method fails to work to large
distances in the presence of very large reddening: stars can no longer
be detected in the J-band.  The detection threshold of the 2MASS
survey can be seen in Fig.~\ref{CMDs} as a marked absence of stars in
the lower right-hand side.

\paragraph{1E~1841$-$045.}  The reddening shows an enormous jump at
3.5--4\,kpc, associated with the Scutum arm.  The estimated reddening
of the source, however, is larger still, and places it behind this
spiral arm.  We thus can only set a lower limit on its distance,
$d>5\,$kpc.  Fortunately, this AXP is associated with a supernova
remnant, Kes~73, for which the estimated distance of 7\,kpc (derived
from \ion{H}{1} absorption measurements; Vasisht \& Gotthelf, 1997) is
consistent with our analysis. We use the latter value below.

\paragraph{XTE~J1810$-$197.} The reddening shows a large jump at
$\sim\!3$\,kpc, from $A_V=4$ to 13.  Since the AXP's estimated
reddening of $A_V=7.8$ is inside this range, the object is almost
certainly confined to the spiral arm that causes the jump.  While the
reddening estimate is less secure, since it is based on a fit to the
broad-band X-ray spectrum, the jump seen in the run of reddening as a
function of distance is so large that the distance measurement should
be secure.  With this distance, XTE~J1810$-$197 is probably the
closest of the AXPs.  Our estimate is consistent with the results of
Gotthelf et al. (2004), who used the extinction estimated from the
X-ray spectrum to suggest a distance in the range 3--5\,kpc (based on
optical reddening studies), and set a firm upper limit to the distance
of 5\,kpc by comparing with the hydrogen column and distance inferred
from \ion{H}{1} measurements to a nearby (but unassociated) supernova
remnant.
\clearpage
\begin{deluxetable}{lclll}
\tablecaption{Distances to the Galactic Anomalous X-ray 
Pulsars\label{distances} \label{fields}}
\tablewidth{0pt}
\tablehead{
\colhead{Object} & \colhead{Field Radius} & \colhead{$N_H$} &
\colhead{$A_V$} & \colhead{Distance\tablenotemark{a}}\\
&\colhead{(arcmin)}&\colhead{$10^{21}$\,cm$^{-2}$}& \colhead{(mag)}&
\colhead{(kpc)}}
\startdata 
4U~0142+61 & 12.1& 6.4$\pm$0.7 &\phn$3.5\pm0.4$& $3.6\pm0.4$ \\
1E~1048.1$-$5937 &\phn6.3&10.0&\phn$5.6$& $9.0\pm1.7$ \\
1E~2259+589 & 10.6&11.2$\pm$3.3&\phn$6.3\pm1.8$& $7.5\pm1.0$ \\
1RXS~J170849.0$-$400910&\phn5.3&13.8$\pm$4 &\phn$7.7\pm2.2$& $3.8\pm0.5$ \\
1E~1841$-$045 &\phn5.3& 25 &\phn14& $\ge 5$  \\
XTE~J1810$-$197 &\phn6.0&14&\phn7.8& $3.1\pm0.5$ 
\enddata
\tablecomments{The three extinction estimates with uncertainties are
  from measurements of edges in high-resolution X-ray spectra by
  Durant \& Van Kerkwijk (2006).  The  other estimates, taken from
  the compilation of Woods \& Thomson (2004), are inferred from
  broad-band fits to the X-ray spectra.  Consequently, their
  uncertainties are difficult to quantify, though likely they are
  below 10\% (see text).}
\tablenotetext{a}{Uncertainty in the distance is dominated by the
  width of the bins in $K$, about 15\%.}
\end{deluxetable}
\clearpage
\section{Comparison with Previous Work}\label{consistency}

Above, we have taken estimates of the reddening to AXPs and used them
to infer distances by matching them against the run of reddening with
distance inferred from red-clump stars.  Here, we compare these
estimates with results for three AXPs that have been studied in
detail. We also include known reddenings and distances of objects near
the AXPs as test-cases for the red-clump method, particularly near
1E~2259+586. 

\subsection{4U~0142+61}
Hulleman et al. (2004) discussed the various distance estimates for 4U
0142+61, which range from 1\,kpc to 4\,kpc.

The near edge of the Perseus Arm (as defined by its forward shock) is
located at a distance of about 3--3.5\,kpc in this direction (e.g.,
Cordes \& Lazio 2002, and references therein), with dense material and
stars within a few 100\,pc of this.  Our distance estimate of
$3.8\pm0.4\,$kpc is fully consistent with the AXP being in this arm.

Our inferred run of reddening with distance is consistent (and greatly
improves upon) what was derived from spectral-energy distribution fits
of field stars by Hulleman et al.\ (2004).  For further verification,
we also obtained low-resolution classification spectra for a number of
brighter stars within a few arcminutes of the source, using the David
Dunlap 1.88\,m telescope at Richmond Hill, Canada.  From the spectral types
and the observed optical magnitudes and colours, we confirm that out
to a distance of $\sim\!3$\,kpc, the reddening does not rise above
$A_V=2$.  Unfortunately, however, our sample did not include any stars
at greater distance and reddening.

\subsection{1E~1048.1$-$5937}
Our distance estimate of $8.9\pm1.9\,$kpc is in stark disagreement
with the estimate of $\sim\!2.7\,$kpc derived by Gaensler et al.\
(2005) from an H{\sc i} bubble that is positionally coincident with
the source and has no other known source.  Our estimate places the AXP
towards the far side of the Carina Arm rather than the near edge.
Gaensler et al.'s distance would be very hard to reconcile with the
large extinction towards this source, since there is little
interstellar dust out to 2.7\,kpc.  We therefore suggest that the
bubble Gaensler et al.\ found is not associated with the AXP.

\begin{figure}
\includegraphics[width=\hsize]{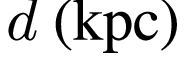}
\caption{Comparison of reddening as a function of distance in the field of 1E
  1048.1$-$5937 from the red-clump method (shown with error bars) with
  other reddening/distance determinations .}\label{morechecks} 
\end{figure}

To check the performance of the red-clump method in this most complex
of fields, we compare the function $A_V(d)$ found above with measured
distances and reddenings of different types of source in the field:
open clusters (from Loktin et al, 1994), Wolf-Rayet stars (van der
Hucht, 2001), O-stars (from Garmany et
al. 1982) and other early-type stars (Guarinos, 1992) within 40\arcmin
of the source. Figure \ref{morechecks} shows all these data against
the reddening curve found for 1E 1048.1$-$5937 above. Although, as
shown in \S\ref{varify}, there is a fair amount of scatter due to
inhomogeneous reddening in the field, the general trend of reddening with
distance is well-followed by all of the data-sets, with the notable
exception of WR~29 (at large distance and low reddening). Crucially,
none of them indicate the presence of 
$A_V>3$ up to 4\,kpc, so the AXP would need a substantially lower
column than that inferred from the X-ray spectrum in order to be at
the distance of Gaensler et al's interstellar bubble.

We also checked the extinction model of Drimmel et al. (2003) to see if
their results were consistent with ours. We found that their
predicted extinction along this line of sight is consistently higher
than ours in the 4--8\,kpc range. This is not surprising, given that
we know from Gaensler et al. that the source lies through a hole in
the interstellar extinction. The Drimmel et al. curve is consistent
with curves 2, 3 and 4 from Figure \ref{circle}. 

\subsection{1E~2259+586}
The supernova remnant CTB~109, thought to be associated with the AXP
1E~2259+586, has been the subject of a number of studies to try to
determine its distance and reddening, with range of previous distance
estimates for the magnetar and SNR as about 3 to 7\,kpc (Kothes et
al. 2002,  introduction). Most recently, Kothes et al.\
(2002) inferred that CTB~109 was at a distance of 3\,kpc, from the
following argument.  The supernova remnant is apparently interacting
with a large molecular cloud in the direction $l\approx108.8\degr$,
$b\approx-0.9\degr$.  This cloud has a velocity of
$-50{\rm\,km\,s^{-1}}$, which is similar to the range of
$-30\ldots-55{\rm\,km\,s^{-1}}$ found for several \ion{H}{2} regions
within a few degrees of the remnant.  For these latter regions,
spectroscopic distance have been measured, which group around a
distance $d=3\,$kpc.  This is rather nearer by than expected based on
the standard rotation curve for this direction, premably because of
streaming motion known to be associated with the Perseus Arm shock.

A distance of 3\,kpc for 1E~2259+586, however, is excluded by our
analysis: at 3kpc, there are no red-clump stars sufficiently highly
reddened to be consistent with 1E~2259+589's column.  One reason for
the discrepancy might be that 1E~2259+589 is not associated with the
SNR CTB~109.  We believe this is unlikely, however, given both the
positional coincidence and the morphology.  Similarly, the shape of the
remnant strongly suggests that it is indeed interacting with a dense
molecular cloud positioned at the above mentioned co-ordinates.

Of the \ion{H}{2} regions listed in Table 2 and shown in Figure 2 of
Kothes et al.\ (2002), two have distances consistent with being behind
the Perseus shock (Sh~149 and Sh~156, $d=4\ldots8\,$kpc).
Interestingly, among the 11 regions with velocities
similar to the cloud with which CTB~109 appears to be interacting
($v<-40{\rm\,km\,s^{-1}}$), these two are amongst the three closest to
CTB~109 on the sky.

Figures~4 and 5 of Kothes et al.\ (2002) show CO and \ion{H}{1} data
in the direction of CTB~109.  From these figures, there appear to be
two components to the CO cloud West of the SNR: one which includes the
``arm'' which passes across the SNR (with velocity
$-47\ldots-51{\rm\,km\,s^{-1}}$) and a second which does not
($-53\ldots-56{\rm\,km\,s^{-1}}$) . These two components show
different morphology and span a large range in velocity, suggesting
that they are physically distinct.  The \ion{H}{1} data is less clear,
but the morphology at $-55{\rm\,km\,s^{-1}}$ again seems match what
one would expect from interaction with the supernova remnant.  Since
the remnant is known not to be interacting with the ``arm'' (see
Sasaki et al.\ 2004), the above-described morphology suggests it is
likely interacting with the second, more distant cloud.

A resolution to the distance discrepancy would thus be that the
foreground cloud, including the ``arm'' which partially covers CTB
109,  is indeed involved in streaming motion, but that
CTB~109 is interacting with a background cloud unconnected with the
first.  If this background cloud is not associated with the streaming
motion, but follows the standard rotation curve
($\Theta_\odot=220{\rm\,km\,s^{-1}}$, $R_\odot=8.5\,$kpc), then its
distance is $\sim\!6\,$kpc, which is consistent with our estimate for
the distance of 1E~2259+589.

Finally, as an independent check, we note that the Wolf-Rayet star
WR~158, which is at an angular separation of 6.1\degr, has a
spectroscopic distance of $\sim\!8\,$kpc and extinction $A_V\approx4$
(van der Hucht, 2001).  Another, WR~151, at a separation 6.8\degr, has
$d\simeq5.7\,$kpc and $A_V\approx3.8$.  These combinations of
reddening and distance are consistent with our inferred run of
reddening with distance (Fig.~\ref{output}).  Thus, we believe our
distance estimate is reliable.

\begin{figure}
\includegraphics[width=\hsize]{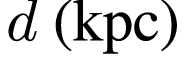}
\caption{The red-clump method as applied to Cas A. The vertical dashed
line is the estimated distance of Cas A, and the grey shaded region
contains the majority of reddening estimates for the SNR and the
central source (a range of estimates rather than the uncertainty on one
estimate).}\label{morechecks2}
\end{figure}

We check further that the red-clump method gives reasonable distances
for objects of known reddening and distance in this part of the
sky. The first object we check is Cas A, which is similar in some
respects to CTB~109: it is a supernova remnant with a central compact
X-ray source. The distance to Cas A is well known at
$d=3.4^{+0.3}_{-0.1}$\,kpc (Reed et al. 1995).
Figure \ref{morechecks2} shows the
derived function $A_V(d)$ for 2000 stars within 10\arcmin of the
position of Cas A. The reddening of the supernova remnant has been
estimated from optical spectra and colors of various sections. The
range of estimates of reddening is shown, which includes the majority
of the estimates and the poorly measured hydrogen column for the
central X-ray source (Fesen \& Hurford, 1996; Fesen et
al. 2006). (Some estimates lie outside this region, but we discount
these extreme estimates). Clearly, the measured distance to the SNR
is consistent with the reddening: our method gives a limit on the
distance $d>1.9$\,kpc. Were we trying to find the distance to Cas A,
we would have reasoned that it must reside in the spiral arm (since it
is also the product of a short-lived, massive star), close to its
known distance  (shown by the vertical dashed line).

\section{Implications}

With our distances, X-ray luminosities for the AXPs 
\footnote{The energy associated with the SNRs also depends on
  distance. For CTB 109, Sasaki et al. (2004) found an explosion energy of
  $0.7\times10^{51}(d/3$\,kpc)$^{2.5}$\,erg, or
  6 -- 8$\times10^{51}$\,erg for our distance. This makes it a
  more energetic supernova than most, suggesting that perhaps AXPs
  generally have higher than typical supernova energies, in contrast
  to what was inferred previously (Vink, 2006; 2006,
  pers. comm.).}
can be calculated
from their unabsorbed flux (taken when they were not thought to be in
outburst; we use the numbers from Woods \& Thomson, 2004).  The
results are listed in Table~\ref{pulsed}.  For comparison, we also
include two sources not considered above, CXOU~J010043.1$-$721134 and
SGR~0526$-$66, for which reasonably good distances are known since
they are located in the SMC and LMC, respectively.  Of these, the
first is thought to be an AXP (Lamb et al.\ 2002; Majid et al.\ 2004)
and the latter is a Soft Gamma-Ray Repeater (SGR), another class of
magnetar.

From Table~\ref{pulsed}, one sees that all Galactic AXPs have
remarkably similar X-ray luminosities.  In part, this may be a
coincidence: all sources are known to vary to some degree, and
XTE~J1810$-$197 is a transient.  Furthermore, while for SGR 0526$-$66,
which is clearly brighter, one might appeal to it being a different
type of source (see also below), the one extra-galactic AXP, CXOU
J010043.1$-$721134 in the SMC, is substantially fainter.
Nevertheless, taken together, the luminosities listed in Table
\ref{pulsed} suggests that the emission mechanism responsible for
the persistent soft X-rays is self-limiting in the AXPs, despite
differences in their spectra and timing properties.

This result was predicted.  Thompson \& Duncan (1996, \S3.6) noted
that the soft X-ray luminosity of a magnetar must saturate at a value
near $L_X\simeq10^{35}{\rm\,erg\,s^{-1}}$, because at higher
luminosities neutrino emission from the interior would become dominant
and cause rapid cooling.  Here, the precise limiting luminosity could
be greater by a factor of a few, depending on the surface composition
(the above limit is for iron, C.~Thompson 2006, pers.\ comm.).
Furthermore, Arras et al.\ (2002) found, from considerations of the
amibipolar diffusion time-scale of the magnetar's internal magnetic
field, that AXPs should remain near the limit for $10^3\dots10^5\,$yr
(i.e., the typical characteristic ages of the AXPs, Woods \& Thompson
2004) before exhausting their internal heat supply and cooling rapidly
by photon emission from the surface.  Thus, it appears that the tight
clustering in $L_X$ we find has a natural explanation in the context
of the magnetar scenario. For a transient AXP, it would appear that
the general luminosity is below the critical value, but that in
outburst it is also limited by the threshold.

\clearpage
\begin{deluxetable}{lcccccc}
\tablecaption{Inferred quantities \label{pulsed}}
\tabletypesize{\footnotesize}
\tablewidth{0pt}
\tablehead{
\colhead{Object} & \colhead{$L_X(2-10{\rm\,keV})$} &
\colhead{$\frac{L_{\rm bb}^{\rm bol}}{L_X}$} &\colhead{$B_{\rm dipole}$} &
\colhead{$T_{\rm bb}$} &
\colhead{$R_{\rm bb}$} & \colhead{Pulsed Fraction}\\
 & \colhead{($10^{35}{\rm\,erg\,s^{-1}}$)} & &\colhead{($10^{14}$\,G)} & 
\colhead{(keV)} & \colhead{(km)} & \colhead{(\%RMS)}} 
\startdata 
4U~0142+61 & 1.3& 0.43 & 1.3 & 0.39 & 8.9 & \phn3.9 \\
1E~1048.1$-$5937 & 1.4 & 0.13 & 3.9 & 0.63 & 1.7 & 62.4 \\
1E~2259+589 & 1.3 & 0.11 &0.6 & 0.41 & 4.2 & 23.2 \\
1RXS~J170849.0$-$400910 & 1.1 &0.27& 4.7 & 0.44 & 5.0 & 20.5 \\
1E~1841$-$045\tablenotemark{a}& 1.1 &0.32& 7.1 & 0.44 & 5.5 & 13\phd\phn \\
XTE~J1810$-$197\tablenotemark{b} & 1.3 & 0.56& \phn2.9 &
0.67 & 2.6 & 42.8\\ 
[2ex]
CXOU~J010043.1$-$721134 & 0.4\tablenotemark{c} &
\nodata\tablenotemark{c} &4.5 & 0.41\tablenotemark{c} & $<\!7$\tablenotemark{c} & 33\phd\phn \\
SGR~0526$-$66 & 2.6 &0.06& 7.4 & 0.53 & 2.6 & \phn4.8
\enddata
\tablecomments{The dipole magnetic field strength $B_{\rm dipole}$,
  black-body temperature $T_{\rm bb}$ and pulsed fraction are taken
  from the compilation of Woods \& Thompson (2004).  The black-body
  radius $R_{\rm bb}$ is calculated using 
  the following fits: 0142, White et al.\ (1996); 1048, Mereghetti et
  al.\ (2004); 2259, Woods et 
  al.\ (2004); 1708, Rea et al.\ (2003); 1841, Morii et al.\ (2003);
  1810, Gotthelf et al.\ (2004); 0100, Lamb et al.\ (2002); 0526,
  Kulkarni et al.\ (2003).  We do not list uncertainties, since these
  are dominated by source variability.}
\tablenotetext{a}{We use the distance $d=7\,$kpc derived by Sanbonmatsu
  \& Halfand (1992); it is consistent with our lower limit of 5\,kpc.}
\tablenotetext{b}{From the spectrum taken soon after outburst; see
  Gotthelf et al.\ (2004).} 
\tablenotetext{c}{The numbers are inferred using the black-body only
  fit by Lamb et al.\ (2002).  As a result, the temperature should be
  considered somewhat uncertain and the black-body radius is an upper
  limit.  The spectrum can also be reproduced adequately with a power
  law (Majid et al. 2004); this yields the same 2--10\,keV
  luminosity within about 20\%.} 
\end{deluxetable}
\clearpage
Typically, the soft X-ray spectra of the AXPs are fit with a composite
model, consisting of a black-body and a power-law component.
Presumably, the former arises from the neutron-star surface, and
typically is responsible for the peak in the observed spectra.  With
our new distances, we can estimate the emitting area, or,
equivalently, the effective black-body radius.  This estimate should
be fairly robust, since in essence it depends only on the wavelength
and the measured flux of the spectral peak.  

In Table~\ref{pulsed}, we list the inferred black-body radii for each
of the AXPs, as well as a number of other physical parameters which
one might expect to be related: the fraction of the total luminosity
that is in the black-body component, the black-body temperature, the
magnetic field strength inferred from the spin-down rate (assuming
magnetic-dipole radiation), and the pulsed flux.  All numbers were
taken when the objects were thought to be in quiescence (with the
exception of XTE~J1810$-$197, which is a transient AXP).  

From the Table, one sees that there is no obvious correlation with
dipole magnetic field strenght, but the black-body radius and
temperature are correlated, with temperature increasing for smaller
effective radii.  The strongest correlation, however, is between the
pulsed fraction and black-body radius: they are well described by
${\rm PF}\propto R_{\rm bb}^{-2}$.  This suggests an inverse
proportionality of pulsed fraction with emitting area, as might be
expected if different sources had, e.g., different sizes of the
regions near the magnetic poles, where heat conduction is largest.
(Such a correlation should break down for sources with hot poles close
to the rotation axis, or the rotation axis close to the line of sight.
But for random mutual inclinations, such geometries are relatively
rare, so it is not surprising that our list does not include such an
exception.)  Our result also implies that any high-energy power-law
component within the 2--10\,keV band must be tied strongly to the
thermal photon flux, since otherwise the correlation would be
weakened.

An open question, though, is what causes the differences between the
AXPs.  In general, one might expect hotter, smaller surfaces for
stronger magnetic fields, since stronger fields should lead to larger
differences in heat conduction across and along magnetic field lines.
But we do not find any correlation with the field strength inferred
from the spin-down rate.  One possibility is that the latter is not a
good measure for the field strength at the surface, e.g., because the
surface field is dominated by higher-order multipole components.  For
that case, however, one might expect to see more complicated pulse
shapes than are typically observed.  Another possibility would be that
the magnetic field strengths are roughly dipolar, but offset from the
center (as typically inferred for magnetic white dwarfs and Ap stars;
e.g., Wickramasinghe \& Ferrario 2000).  If so, the AXPs with the
largest pulsed fractions might have the largest offsets, for which the
effects of partial overlap between polar caps might lead to smaller
effective radii.

While the above correlations are intriguing, one sees from
Table~\ref{pulsed} that there is one exception, SGR~0526$-$66.  This
object, although discovered in 1979 as the first SGR through its
gamma-ray outbursts, is the most AXP-like of its class in its timing
and spectral properties (Kulkarni et al.\ 2003).  Yet, it flouts both
the temperature and black-body radius correlations with pulsed
fraction, and has a higher 2--10\,keV luminosity than the limit
inferred for the AXPs.  The luminosity of the black-body component
alone, however, is in the same range as for the AXPs.  If the emission
source for the thermal component is the same as it is for the AXPs,
this suggests that the non-thermal component is powered at least
partly from another, additional reservoir of energy.  For instance,
the energy could be stored in a large magnetic twist imparted during
the previous gamma-ray bursting phase.  Such magnetic twists are
thought to decay on time scales of the order decades (Beloborodov \&
Thompson 2006), so it is possible that SGR~0526$-$66 has not yet
returned to a quiescent, AXP-like state.

Intriguingly, it would seem that the AXPs, particularly 1E
1048.1$-$5937 and 1E 2259+586 seem to be located through gaps in the
interstellar extinction. This likely is selection effect: the objects
are detected primeraly in the soft X-ray band due to the sensitivities
of the main X-ray observatories, and the X-ray spectra of the AXPs
fall rapidly in the 2--10\,keV range. The only two AXPs with
estimated hydrogen columns $N_H>1.5\times10^{22}$\,cm$^{-2}$, 1E
1841$-$045 and AX J1845$-$0258 were discovered through observations of
the SNRs Kes~73 and Kes~75 respectively (Kriss et al. 1985; Gotthelf
\& Vasisht, 1998). This suggests that there may be many more AXPs in
the Galaxy, which have not been discovered so far due to high
extinction. They could, however, be discovered by their high-energy
($>$20\,keV) emission, if it exists and is pulsed, as has been found for
the AXPs detected by INTEGRAL so far (Kuiper et al. 2006).

\section{Conclusions}

We have applied the ``red-clump'' method to 2MASS data to construct
the run of reddening versus distance in the directions of each of the
Galactic AXPs.  Combined with estimates of the reddenings to the AXPs,
half of which are from our recent, model-independent determinations from
photo-electric absorption edges in high-resolution X-ray spectra, we
inferred distances.  We found that two of these estimated distances
are inconsistent with ones in the literature, but found likely reasons
for the discrepancy in both cases and concluded that our results were
reliable.

From the reddening versus distance diagrammes, we find that the AXPs
tend to fall in regions of rapidly rising reddening that are
associated with spiral arms.  This is not surprising, since they are
the young remnants of short-lived massive stars.  In particular,
4U~0142+61 has a position consistent with the Perseus Arm,
1E~1048.1$-$5937 lies on the far side of the Carina Arm, and
1XTE~J1810$-$197 and 1RXS~J170849.0$-$400910 fall along the
Crux-Scutum arm.  1E~1841$-$045 could either lie in the Scutum arm or
in the Molecular Ring, which dominates the gas and dust density at
galactocentric distances of about 4\,kpc (Dame et al.\ 2001).
1E~2258+586 falls near the end of the Outer Arm, known to exist in
this direction beyond the Perseus arm (e.g., Kimeswenger \& Weinberger
1989).

From our distances, we infer 2--10\,keV luminosities and we find that
these cluster tightly around $1.3\times10^{35}{\rm\,erg\,s^{-1}}$,
consistent with the prediction in the context of the magnetar model
that a saturating luminosity must exist above which rapid internal
neutrino cooling is effective.  Furthermore, we calculated effective
emitting radii for the thermal components in the X-ray spectra, and
find that these are inversely correlated with temperature, while the
corresponding areas are inversely proportional to the pulsed fraction.
This suggests the internal heating is released predominantly through
one or more hot polar caps, with sizes that differ between the
different AXPs. 

The red-clump method can be applied to any line of sight in the
Galactic plane, and is particularly useful combined with reddening
estimates from X-ray spectroscopy.  More accurate results may be
obtained using deeper infrared imaging of selected fields, although,
unfortunately, it may not be possible to extend the method to regions
with very high reddening, since the red clump stars may well become
confused with highly reddened main-sequence stars.  As a result of the
latter limitation, the method may not be generally useful for the
other class of magnetars, the soft gamma-ray repeaters, which
generally suffer very high extinction.  It should be useful, however,
for point sources such as the Compact Central Objects.

Generally, for further analysis of distances and structure within the
Milky Way, it would be useful to cross-calibrate results from the
red-clump method with those from X-ray absorption studies, X-ray dust
scattering haloes, and \ion{H}{1} and CO measurements. 

\medskip\noindent{\bf Acknowledgments:} We would like to acknowledge
the thoroughness and many useful suggestions and comments of the
anonymous referee. We thank David Kaplan for
pointing out the paper describing the red-clump method to us, and
Martin L\'opez-Corredoira for information on the applicability of the
red-clump method.  We also thank Bryan Gaensler for help with
interpreting CO and \ion{H}{1} data and for discussions about
1E~1048.1$-$5937, Manami Sasaki and Terrance Gaetz for discussing the
case of CTB~109, and Tom Dame and Peter Martin for general discussions
of Galactic gas and dust. This publication makes use of the Two Micron
All Sky Survey, which is a joint project of the
University of Massachusetts and the Infrared Processing and Analysis
Center/California Institute of Technology, funded by NASA and NSF.  We
acknowledge financial support from NSERC.

\end{document}